\documentclass[aps,twocolumn,preprintnumbers,superscriptaddress,amsmath,amssymb,footinbib,prl]{revtex4}
\usepackage{epsfig}
\usepackage{graphicx}
\usepackage{bm}
\newcommand{\be}{\begin{equation}}
\newcommand{\ee}{\end{equation}}
\newcommand{\bea}{\begin{eqnarray}}
\newcommand{\eea}{\end{eqnarray}}

\begin{document}
\title{Hydrodynamics with a chiral hadronic equation of state including quark degrees of freedom}
%\author{ J.~Steinheimer,Veronica.., H.~Petersen, S.~Schramm, M.~Bleicher, H.~St\"ocker}
%\affiliation{Institut f\"ur Theoretische Physik,\\
%Johann Wolfgang Goethe-Universit\"at,\\
%Max-von-Laue Str.\ 1,\\
%60438 Frankfurt, Germany}

\author{J.~Steinheimer}
\affiliation{Institut f\"ur Theoretische Physik, Goethe-Universit\"at, Max-von-Laue-Str.~1,
D-60438 Frankfurt am Main, Germany}

\author{V.~Dexheimer}
\affiliation{Frankfurt Institute for Advanced Studies (FIAS), Ruth-Moufang-Str.~1, D-60438 Frankfurt am Main,
Germany}

\author{M.~Bleicher}
\affiliation{Institut f\"ur Theoretische Physik, Goethe-Universit\"at, Max-von-Laue-Str.~1,
D-60438 Frankfurt am Main, Germany}

\author{H.~Petersen}
\affiliation{Institut f\"ur Theoretische Physik, Goethe-Universit\"at, Max-von-Laue-Str.~1,
D-60438 Frankfurt am Main, Germany}

\author{S.~Schramm}
\affiliation{Institut f\"ur Theoretische Physik, Goethe-Universit\"at, Max-von-Laue-Str.~1,
D-60438 Frankfurt am Main, Germany}
\affiliation{Frankfurt Institute for Advanced Studies (FIAS), Ruth-Moufang-Str.~1, D-60438 Frankfurt am Main,
Germany}
\affiliation{Center for Scientific Computing, Max-von-Laue-Str.~1, D-60438 Frankfurt am Main}

\author{H.~St\"ocker}
\affiliation{Institut f\"ur Theoretische Physik, Goethe-Universit\"at, Max-von-Laue-Str.~1,
D-60438 Frankfurt am Main, Germany}
\affiliation{Frankfurt Institute for Advanced Studies (FIAS), Ruth-Moufang-Str.~1, D-60438 Frankfurt am Main,
Germany}
\affiliation{GSI Helmholtzzentrum f\"ur Schwerionenforschung GmbH, Planckstr.~1, D-64291 Darmstadt, Germany}

%\author{D.~Zschiesche}
%\affiliation{Institut f\"ur Theoretische Physik, Johann Wolfgang Goethe-Universit\"at, Max-von-Laue-Str.~1,
%D-60438 Frankfurt am Main, Germany}

\begin{abstract}
We investigate the influence of a deconfinement phase transition on the dynamics of hot and dense nuclear matter.
To this aim a hybrid model with an intermediate hydrodynamic stage for the hot and dense phase of the system is employed for collisions of Pb+Pb/Au+Au at beam energies of $E_{\rm lab}=2-160A~$GeV, while initial
and final interactions are performed by a microscopic transport approach (UrQMD). In the hydrodynamic stage an equation of
state that incorporates a critical end point (CEP) in line with lattice data is used. It follows from coupling the Polyakov loop (as an order parameter for deconfinement) to a chiral hadronic $\rm{SU(3)_f}$ model. In this configuration the EoS describes chiral restoration as well as the deconfinement phase transition. We compare the results from this new equation of state to results obtained, by applying a hadron resonance gas equation of state, focusing on bulk observables deemed to be sensitive to the phase transition to a Quark-Gluon Plasma.
\end{abstract}

\maketitle

\noindent

Heavy ion collisions at intermediate incident beam energies ($E_{\rm lab}=5-200A~$GeV) offer the unique opportunity of being able to scan a wide range of temperatures $T$ and baryo-chemical potentials $\mu_B$ in the phase diagram of strongly interacting matter \cite{Gyulassy:2004zy} (for
recent lattice QCD results see \cite{Fodor:2001pe,Fodor:2007vv,Karsch:2004wd}, for
phenomenological studies see
\cite{Stephanov:1998dy,Gazdzicki:1998vd,Stephanov:1999zu,Bravina:1999dh,Bravina:2000dk,Gazdzicki:2004ef,Arsene:2006vf}).
In this energy region one hopes to find experimental evidence for a deconfinement phase transition from hadronic matter to the Quark-Gluon Plasma (QGP) phase (where quarks are deconfined).
Especially the so-called critical end point (CEP), a point in the phase diagram that terminates the first order
phase transition-line (which is expected for high chemical
potentials), is of great interest.\\

Key bulk observables like the directed flow $v_1$
\cite{Ollitrault:1992bk,Rischke:1996nq,Sorge:1996pc,Heiselberg:1998es,Soff:1999yg,Brachmann:1999xt,Csernai:1999nf,Zhang:1999rs,Kolb:2000sd,Bleicher:2000sx,Stoecker:2004qu,Zhu:2005qa,Petersen:2006vm}, but also particle multiplicities, ratios, and their fluctuations, have been predicted and sometimes already shown to be sensitive to the active degrees of
freedom in the early stage of the reaction. Indeed, the energy dependences of various
observables show anomalies at low SPS energies which might be related to the onset of
deconfinement and chiral symmetry restoration \cite{Gazdzicki:2004ef,Gazdzicki:1998vd}.

Early-on, fluid dynamics has been proposed as an elegant way to include the EoS of strongly interacting matter in the description of heavy ion collisions \cite{Hofmann:1976dy,Stoecker:1986ci,Hung:1994eq}.
Especially since experiments at the RHIC facility have claimed to have found a (s)QGP that behaves like a nearly ideal fluid, the idea of modeling heavy ion collisions with (ideal) fluid dynamics has been revived \cite{Kolb:2003dz,Baier:2006gy,Song:2007fn}.

In order to study the fluid dynamical evolution of a heavy ion collision, the different boundary conditions (i.e. the initial space-time distributions of the corresponding energy and baryon density, as well as the freeze out prescription) have to be determined.
Since experimental data provide mainly information from the final state of the reaction, integrated over the time evolution of the system, the initial
state for hydrodynamical simulations is usually inferred from model assumptions or by an 'educated guess' in
comparison to data. The connection between (observed)
final state and the inferred initial conditions is further blurred by the unknown equation of state, potential viscosity effects,
and problems in the freeze-out treatment. Another issue concerns the assumption of thermal equilibrium, which is probably not fulfilled for the early stages of heavy ion collisions at intermediate energies.

There have been attempts to solve these problems by describing such collisions with viscous or multi-fluid-hydrodynamic
models
\cite{Mishustin:1988mj,Katscher:1993xs,Brachmann:1997bq,Bleicher:1998xi,Russkikh:2003ma,Ivanov:2005yw,Toneev:2005yy},
but the practical application of these models is difficult.

Transport theory offers another, different, approach aiming at the consistent description of heavy-ion reactions, from the initial state to the final decoupling of the system. This microscopic description has been applied quite successfully to the partonic as well as to the hadronic
stage of heavy ion collisions \cite{Molnar:2004yh,Xu:2004mz,Burau:2004ev}. However, to explain hadronization and the phase transition between the hadronic and the partonic phase on a microscopic
level is one of the main issues to be resolved. It is therefore difficult to find an appropriate prescription of
the phase transition in such a microscopic approach.

To obtain a more comprehensive picture of the whole dynamics of heavy ion reactions various so called micro+macro hybrid approaches have been developed
during the last years \cite{Magas:2001mr}. The NEXSpheRIO approach uses initial conditions that are calculated in a non equilibrium model (NEXUS) followed by an ideal hydrodynamic evolution \cite{Paiva:1996nv,Aguiar:2001ac,Socolowski:2004hw}.
For the freeze-out a continuous emission scenario or a standard Cooper-Frye calculation is employed.
Other groups, e.g. Hirano et al \cite{Hirano:2005xf,Hirano:2007ei,Werner:2009fa} , Bass/Nonaka \cite{Bass:1999tu,Nonaka:2006yn,Dumitru:1999sf}, use
smooth Glauber or CGC initial conditions followed by a full three-dimensional hydrodynamic evolution and
calculate the freeze-out with a subsequent hadronic cascade. The separation of chemical and kinetic freeze-out and final state interactions like resonance decays and rescatterings are taken into account.\\

In this paper, we apply a transport calculation with an embedded three-dimensional ideal relativistic one-fluid
calculation for the hot and dense stage of the reaction, thus reducing the parameters
for both the initial conditions and the freeze-out prescription. This will allow us to compare calculations with different EoS within the same framework and to extract the effect of changes in the EoS  - e.g. a phase transition from hadronic matter to the QGP - on observables. In \cite{Petersen:2008dd} this model has been used, applying a hadron resonance gas EoS, to provide a baseline calculation, disentangling the effects of the different assumptions for the underlying dynamics in a transport vs. hydrodynamic calculation.
We will extend this purely hadronic calculation by introducing an EoS that includes a deconfinement phase transition. The calculations will be performed in the broad energy range from $E_{\rm lab}=2-160A~$GeV where experimental data from BNL-AGS and CERN-SPS exists and which will be explored in more detailed energy scans by the FAIR project near GSI and the RHIC low energy program.

%--------------------------------------------------------------------------

\section{The hybrid model}
The Ultra-relativistic Quantum Molecular Dynamics Model \cite{Bass:1998ca,Bleicher:1999xi} (in its cascade mode) is used to calculate the initial state of a heavy ion collision for the hydrodynamical evolution \cite{Steinheimer:2007iy}. This is done to account for the non-equilibrium dynamics in the very early stage of the collision. In this configuration the effect of event-by-event fluctuations of the initial state is naturally included. The coupling between the UrQMD initial state and the hydrodynamical evolution happens at a time $t_{\rm start}$ when the two Lorentz-contracted nuclei have passed through each other:

\begin{equation}
t_{\rm start}=2R/\sqrt{\gamma_{\rm{c.m.}}^2-1}~~,
\end{equation}

where $R$ is the radius of the lead nucleus and $\gamma_{\rm{c.m.}}$ the Lorenz gamma factor of the two colliding nuclei in their center of mass frame. At this start time all initial collisions have proceeded, i.e. also the initial baryon currents have decoupled from each other, and it is the earliest time at which local thermodynamical equilibrium may be achieved.
To map all 'point-like' particles from UrQMD onto the spatial grid of the hydrodynamic model each hadron is represented by a Gaussian of finite width. This procedure is necessary, since the cell length of $0.2$ fm is much smaller than the actual size of a hadron. The Gaussian width is chosen to be $\sigma= 1$ fm. This width reflects the typical size of hadrons and avoids  numerical instabilities (numerical entropy production) in the initial phase of the hydrodynamical evolution. This instantaneous thermalization at $t_{\rm start}$ goes along with an increase in entropy, as entropy is maximized in the equilibrium state. However, note that is has been checked that the results only weakly depend on the choice of the time $t_{\rm start}$ \cite{Petersen:2008dd}. Especially final particle multiplicities and their mean transverse mass only change by about $10 \%$, if the time $t_{\rm start}$ is doubled. 

For calculations at finite impact parameter, the spectators - particles that have not interacted until $t_{start}$ - are propagated separately from the hydrodynamic evolution. They are treated as free streaming particles until the end of the hydrodynamic phase has been reached.

The full (3+1) dimensional ideal hydrodynamic evolution is performed using the SHASTA algorithm \cite{Rischke:1995ir,Rischke:1995mt}.
The partial differential equations are solved on a three-dimensional spatial Eulerian grid with fixed position and size in the
computational frame. The size of the grid is 200 cells in each direction, while the cell size has been chosen to be $dx=0.2$ fm which leads to time steps of $dt=0.08$ fm in order to avoid non-causal effects in the propagation (Courant criterion).

\begin{figure}[t]
\centering
\includegraphics[width=0.5\textwidth]{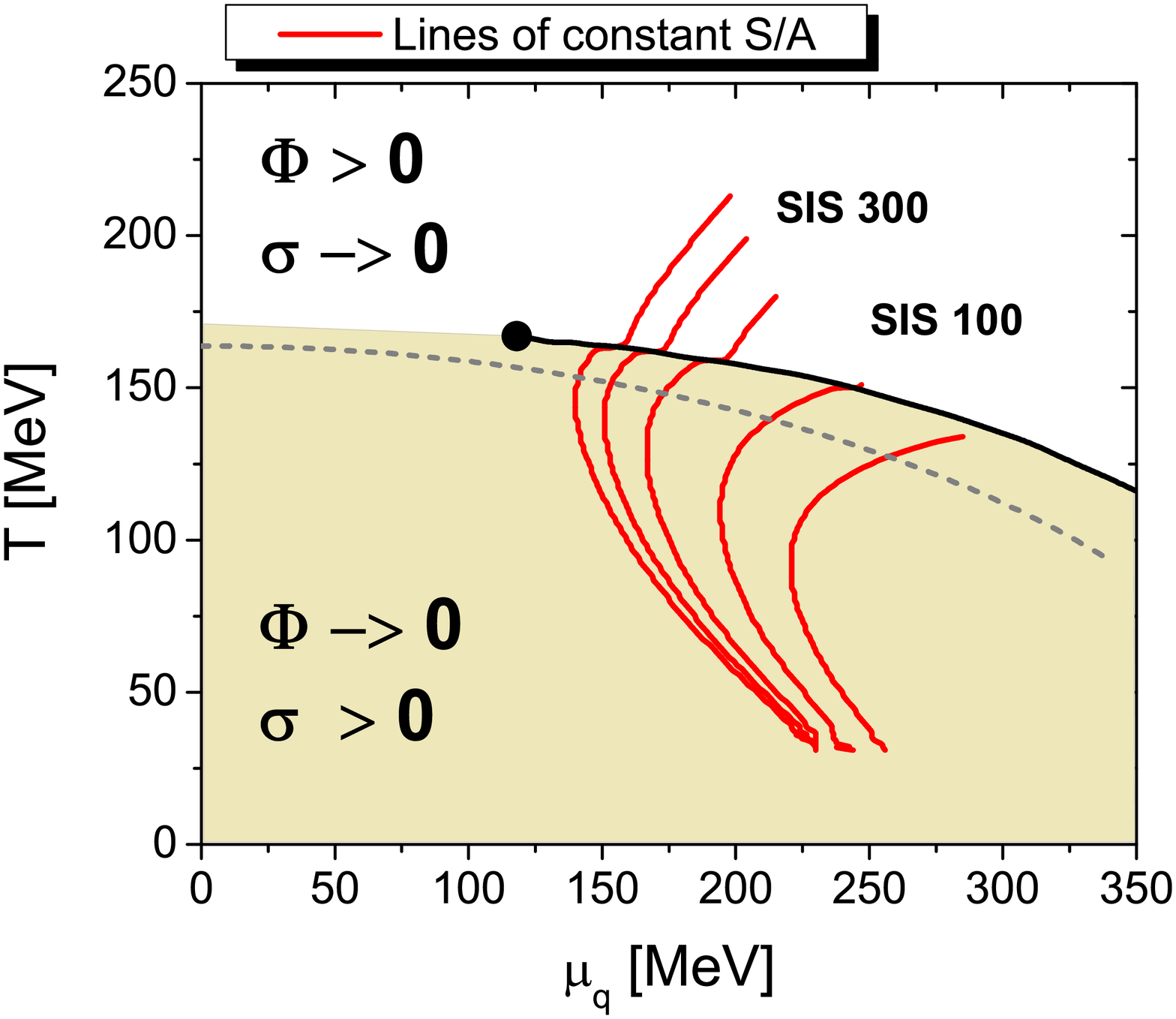}
\caption{\label{tmu}
(color online) Isentropic expansion paths (red lines) in the $T-\mu_q$ plane for very central Pb+Pb/Au+Au reactions.
Isentropic expansion from the overlap model initial conditions are shown as full line in blue. Beam energies are from left to right: $E_{\rm lab}=40, 30, 20, 10, 5 A$~GeV. The line of the first order phase transition is indicated in black together with the critical endpoint of the model. Also shown is the line of constant energy density $\epsilon= 4 \epsilon_0$ (gray dashed).}
\end{figure}

To transfer all particles back into the UrQMD model, an approximate iso-eigentime transition is chosen (see \cite{Li:2008qm} for details). Here, we 'freeze out' individual transverse slices, of thickness $\Delta z = 0.2 $fm, at a constant time-like transition hypersurface. This time for each slice is given, whenever the energy density 
$\varepsilon$, in every cell of this slice, has dropped below four times
the ground state energy density (i.e. $\sim 580 {\rm
MeV/fm}^3$). This assures that all cells have passed through the mixed phase of the equation of state and the effective degrees of freedom at the transition are hadronic (see Fig. \ref{tmu} for a depiction of this line of constant energy density in the $T$-$\mu_q$ phase diagram).\\
By applying a gradual transition one obtains an almost rapidity independent switching temperature. The hydrodynamic fields, in a given slice, are transformed to particle degrees of freedom via the Cooper-Frye equation on an isochronous time-like hypersurface in the computational frame (the hypersurface normal is $d\sigma_{\mu}=(d^3x,0,0,0)$).\\
As different longitudinal slices have different freeze out times, $d\sigma_{\mu}$ should of course also have a space like component. Such a parametrization of the hypersurface is not easily dealt with numerically (especially since our system has locally fluctuating densities and therefore an inhomogeneous hypersurface). We therefore try to justify our approach by comparing freeze out results from a simple, analytically solvable, one-dimensional Bjorken scenario by using both, our approximate and the correct parametrization of $d\sigma_{\mu}$. In a realistic set up, where the longitudinal expansion of the system is about $10$ fm, we obtain a total difference in particle production of about $10 \%$. The error in particle production per rapidity interval grows for larger rapidities. In consequence, we expect the calculated rapidity distributions, of particles produced at the highest SPS energies ($E_{\rm lab}=160A~$GeV), to show the largest effect of our choice of $d\sigma_{\mu}$. More precisely, we expect the present rapidity distributions at the highest energies to be lower at mid-rapidity and broader at high rapidities, as compared to results with the correct $d\sigma_{\mu}$. As a remark, it is possible to numerically extract the correct parametrization of the full hypersurface using digital image processing techniques \cite{schlei}, and then compare these results with data acquired with our simplified hypersurface. There is work in progress on this task, which will be subject of future publications.\\ 

As has been pointed out in \cite{Petersen:2008dd}, the present transition procedure conserves the baryon number, the electric charge and the total net strangeness on an event-by-event basis, but the total energy only on average and only if the freeze out is treated properly. In the present calculations, the total energy of the system varies by $1-2 \%$ per event. To overcome this problem it is possible to rescale the momenta of all produced particles in every single event to enforce exact energy conservation. We have employed both methods and checked that the results, for particle rapidity and momentum spectra, do not depend on the method used but are identical, within statistical errors, when averaged over 500 events.\\

After the particles are created according to our prescription, they proceed in their evolution in the hadronic cascade (UrQMD) where rescatterings and final decays are calculated until all interactions cease and the system decouples.

A more detailed description of the hybrid model
including parameter tests and results can be found in \cite{Petersen:2008dd}.

\begin{figure}[t]
\centering
\includegraphics[width=0.5\textwidth]{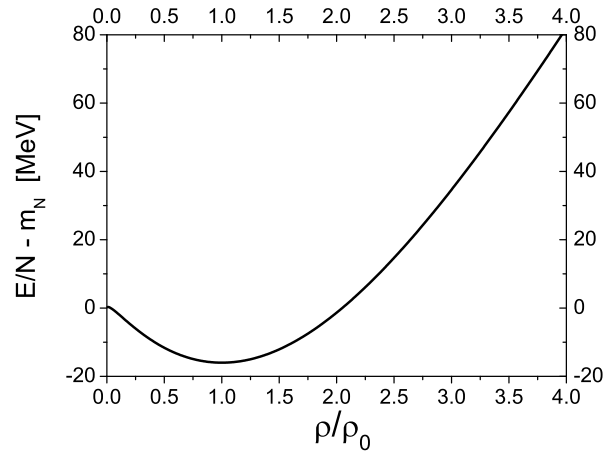}
\caption{\label{binding}
Binding energy per nucleon as a function of the net baryon number density at $T=0$. The minimum at nuclear ground state density ($\rho_0$) corresponds to a binding energy of $-16$ MeV.}
\end{figure}

\section{The equation of state}

After the introduction of the dynamical framework, we now turn to a detailed discussion of the novel EoS. Here, we employ a single model to obtain the EoS of the hadronic and the quark phase.

%The extension of the model to quark degrees of freedom
%is constructed
The hadronic part of the model is a flavor-SU(3) model, which is an extension of a non-linear representation of a sigma-omega model including the lowest-lying multiplets of baryons and mesons (for the derivation and a detailed discussion of the hadronic part of the model Lagrangian see \cite{Papazoglou:1998vr,Papazoglou:1997uw,Dexheimer:2008ax}).
In spirit similar to the PNJL model \cite{Fukushima:2003fw} it includes the Polyakov loop $\Phi$ as an effective field
and it adds quark degrees of freedom.
The temporal background field $\Phi$ is defined as $\Phi=\frac13$Tr$[\exp{(i\int d\tau A_4)}]$, where $A_4=iA_0$ is the temporal component
of the SU(3) gauge field.

The Lagrangian density of the model in mean field approximations reads:
\begin{eqnarray}
&L = L_{kin}+L_{int}+L_{meson},&
\end{eqnarray}
where besides the kinetic energy term for hadrons and quarks, the terms

\begin{eqnarray}
&L_{int}=-\sum_i \bar{\psi_i}[\gamma_0(g_{i\omega}\omega+g_{i\phi}\phi)+m_i^*]\psi_i,&
\end{eqnarray}
\begin{eqnarray}
&L_{meson}=-\frac{1}{2}(m_\omega^2 \omega^2+m_\phi^2\phi^2)\nonumber&\\
&-g_4\left(\omega^4+\frac{\phi^4}{4}+3\omega^2\phi^2+\frac{4\omega^3\phi}{\sqrt{2}}+\frac{2\omega\phi^3}{\sqrt{2}}\right)\nonumber&\\
&+\frac{1}{2}k_0(\sigma^2+\zeta^2)-k_1(\sigma^2+\zeta^2)^2&\nonumber\\
&-k_2\left(\frac{\sigma^4}{2}+\zeta^4\right)-k_3\sigma^2\zeta&\nonumber\\
&-k_4\ \ \ln{\frac{\sigma^2\zeta}{\sigma_0^2\zeta_0}} + m_\pi^2 f_\pi\sigma&\nonumber\\
&+\left(\sqrt{2}m_k^ 2f_k-\frac{1}{\sqrt{2}}m_\pi^ 2 f_\pi\right)\zeta~,&
\end{eqnarray}

represent the interactions between baryons (and quarks)
and vector and scalar mesons the self interactions of
scalar and vector mesons and an explicit chiral symmetry breaking term.
The index $i$ denotes the baryon octet and the three light quarks. Here, the mesonic condensates (determined in
mean-field approximation) included are
the vector-isoscalars $\omega$ and $\phi$ and
the scalar-isoscalars $\sigma$ and $\zeta$ (strange quark-antiquark state). At this point we neglect the $\rho$-meson contributions as we only discuss isospin-symmetric matter.

The effective masses of the baryons and quarks
are generated by the scalar mesons except for a small explicit
mass term and the term containing the Polyakov field $\Phi$ \cite{Dexheimer:2009hi}:
\begin{eqnarray}
&m_{b}^*=g_{b\sigma}\sigma+g_{b\zeta}\zeta+\delta m_b+g_{b\Phi} \Phi^2,&
\end{eqnarray}
\begin{eqnarray}
&m_{q}^*=g_{q\sigma}\sigma+g_{q\zeta}\zeta+\delta m_q+g_{q\Phi}(1-\Phi).&
\end{eqnarray}
With the increase of temperature/density, the scalar fields decrease
in value, causing the effective masses of the particles to decrease towards chiral symmetry restoration.
The Polyakov loop effectively suppresses baryons at high temperatures/densities and quarks at low temperatures/densities
due to their corresponding mass shifts shown above.\\
Due to meson vector interactions, the baryons obtain an effective chemical potential:
\begin{eqnarray}
&\mu_{b}^*=\mu_{b}-g_{b\omega}\omega-g_{b\phi}\phi,&
\end{eqnarray}

All thermodynamic quantities are derived from the grand canonical potential $\Omega$ by assuming an equilibrated state, where $-p=\frac{\Omega}{V}$ (p is the pressure) is in the absolute minimum of $\Omega$ with respect to all fields. For the region of phase coexistence (at the 1. order phase boundary) we assume a phase mixture. This of course implies, that the expanding system is in chemical equilibrium during the whole evolution.
The grand canonical potential of the model has the form:
\begin{equation}
	\frac{\Omega}{V}=-L_{meson}+\frac{\Omega_{th}}{V}-U
\end{equation}
Here $\Omega_{th}$ includes the heat bath of hadronic and quark quasiparticles (as function of $T$, $\mu^*$ and $m^*$)
within the grand canonical potential of the system. The Polyakov-loop potential $U$ will be discussed in the following.

The potential $U$ for the Polyakov loop reads:
\begin{eqnarray}
&U=(a_0T^4+a_1\mu_{B}^4+a_2T^2\mu_{B}^2)\Phi^2&\nonumber\\&+a_3T_0^4\ln{(1-6\Phi^2+8\Phi^3-3\Phi^4)}.&
\end{eqnarray}
It is based on \cite{Ratti:2005jh,Roessner:2006xn} and fitted to the pressure and Polyakov loop values as computed in lattice-QCD calculations at zero chemical potential as discussed in detail in \cite{Dexheimer:2009va}. Additional terms, depending on the chemical potential, are fixed in order to reproduce the phase diagram at high densities. This includes a first order phase transition line in $\mu_q$ and $T$ that ends in a critical point, of second order, at the values obtained by lattice calculations \cite{Fodor:2004nz}.  The coupling constants for the baryons (already shown in \cite{Dexheimer:2008ax}) are chosen to reproduce the vacuum masses of the baryons and mesons, nuclear saturation properties as well as the hyperon potentials. The vacuum expectation values of the scalar mesons are constrained by
reproducing the pion and kaon decay constants.
The coupling constants for the quarks
($g_{q\sigma}=-3.0$, $g_{s\zeta}=-3.0$,
$T_0=200$ MeV, $a_0=1.85$, $a_1=1.44$x$10^{-3}$, $a_2=0.08$, $a_3=0.40$, $g_{N\Phi}=1500.00$ MeV, $g_{q\Phi}=500$ MeV) are
chosen to reproduce lattice data (for $T_0=270$ and pure gauge a first order phase transition at $\mu=0$ and $T=270$ MeV is reproduced) and known information about the phase diagram.

As can be seen in Fig. \ref{tmu} the transition from hadronic to quark matter obtained is a crossover for small chemical potentials.
At vanishing chemical potential the transition temperature is $171$ MeV, determined as the peak of the change of the scalar field
and the Polyakov loop. Beyond the critical end-point (at $\mu_{c,B}=354$ MeV, $T_c=167$ MeV for symmetric
matter in accordance with \cite{Fodor:2004nz}) a first order transition line begins.
As can be seen in Fig. \ref{binding}, the model reproduces nuclear matter saturation at realistic values for the saturation density, nuclear binding energy,
as well as compressibility and asymmetry energy. In addition, realistic results at low
densities for the nuclear matter liquid-gas phase transition are obtained and the model has been successfully applied in order to model the properties of compact stars \cite{Dexheimer:2009hi}.
It is crucial for the numerical studies of heavy-ion simulations to have an equation of state at hand that shows reasonable
behavior over a large range of densities and temperatures, which is the case here.
In the following we will refer to this EoS as the deconfinement EoS (DE).

We will compare results obtained with the DE to calculations using an EoS consisting of a
hadron resonance gas including all reliably known resonances with masses up to $2$ GeV (referred to as hadron gas HG).
The HG is a very important ingredient of the model, because the active degrees of freedom on both sides of the transition hypersurface have to be equivalent to ensure the conservation of important quantities (e.g. entropy). For the HG this is the case, as it has the same degrees of freedom as the UrQMD model. In the DE the hadrons acquire effective masses due to interaction and therefore this equivalence condition is only approximately fulfilled. To solve this problem we change the active equation of state after the last step of the hydrodynamical evolution (from the deconfinement EoS to the HG), thus obtaining the correct temperatures and chemical potentials for the particle distributions. If we use the DE EoS for the freeze-out prescription we find that the total energy conservation is violated systematically by about $3 \%$.
Therefore results on particle spectra also change slightly. As this change is only on the level of a few percent, we prefer to have the correct degrees of freedom during the transition from hydrodynamics to transport.

\begin{figure}[t]
\centering
\includegraphics[width=0.5\textwidth]{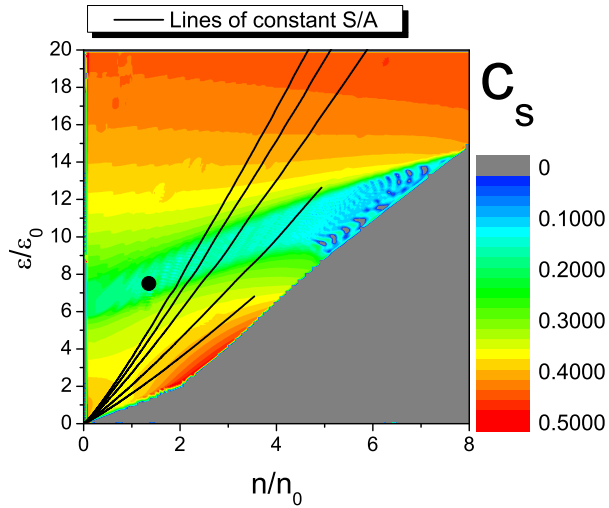}
\caption{\label{erho}
(color online) Isentropic expansion paths (black lines) and contours of the speed of sound in the $\epsilon$-$n$ phase diagram. The beam energies, associated with the lines, are the same as in the previous figure. The gray region relates to unphysical combinations of $\epsilon$ and $n$ ($T\le0$). The critical endpoint of the model is displayed as the black dot.}
\end{figure}

Fig. \ref{tmu} depicts the phase structure of the obtained EoS. Included are lines of constant entropy per baryon as they are expected for beam energies of $E_{\rm lab}=40, 30, 20, 10, 5 A$~GeV (from left to right). The values for $\frac{S}{A}$ were calculated using a simple overlap model where: $n_b = 2 \gamma_{\rm{c.m.}} n_0$, \ $\epsilon = \sqrt{s} \gamma_{\rm{c.m.}}  n_0$, and $n_0$ is the nuclear ground state baryon number density.
It is obvious that the deconfined phase is already reached at energies above $E_{\rm lab}=10 A$ GeV. As has been shown in \cite{Steinheimer:2007iy}, an incident beam energy of $E_{\rm lab}=40 A$ GeV may not be sufficient to reach the critical endpoint suggested in \cite{Fodor:2004nz}. If instead the CEP is situated at higher $\mu_B>450$ MeV (as suggested  by recent lattice studies \cite{deForcrand:2007rq}) it could again be accessible in the energy range of the FAIR project.

\section{The speed of sound}

An important property of a hot and dense nuclear medium is the speed of sound ($c_s$):
\begin{equation}
c_s^2 = \left. \frac{d p}{d \epsilon}\right|_{S/A} = \left. \frac{d p}{d e}\right|_{n}+ \frac{n}{\epsilon + p} \left. \frac{d p}{d n}\right|_{\epsilon}.
\end{equation}
It is not only closely related to expansion dynamics but also controls the way perturbations (sound waves) travel
through the fireball.
Fig. \ref{erho} shows $c_s$ as a contour plot in the
$\epsilon-n$ phase diagram, where $\epsilon$ is the energy density and $n$ the net baryon number density, both given in units of the ground state values ($\epsilon_0 = 146 ~\rm{ MeV/fm^3}$, $n_0 = 0.159 ~\rm{ fm^{-3}}$).

\begin{figure}[t]
\centering
\includegraphics[width=0.5\textwidth]{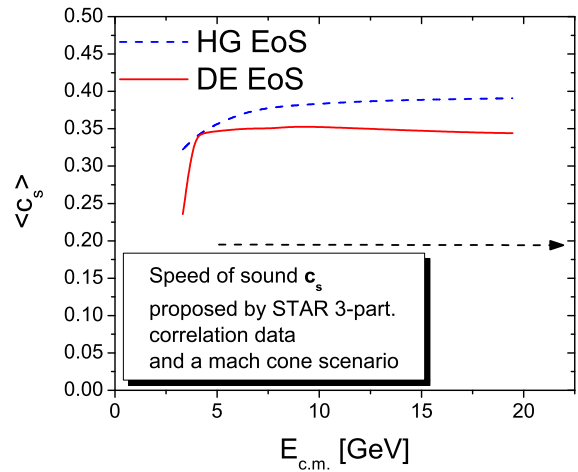}
\caption{\label{cs}
(color online) Excitation function of the averaged speed of sound in most central A+A collisions. The deconfinement EoS (red solid line) is compared to the hadron resonance gas (blue dashed line). Indicated is also the speed of sound extracted from 3 particle correlation studies at the STAR experiment \cite{:2008nd}.}
\end{figure}

One can clearly see the reduction of the speed of sound (softening) of the EoS in the mixed phase. Also depicted is the location of the CEP and isentropic paths for energies as in Fig. \ref{tmu}. Please note that the numerical accuracy for Fig. 2 is very limited when 
going to very low temperatures. This is essentially because the energy
density is proportional to $T^4$. The $T=0$ line can only be thought of as 
a guide for the eye (the error in $T$ is about $30$ MeV for the
lowest value of $\epsilon/\epsilon_0$). In fact we have checked that the binding energy and compressibility 
is indeed well reproduced at nuclear saturation density \cite{Dexheimer:2009va}. In order to show that the
system reproduces the correct nuclear groundstate, Fig. \ref{binding} depicts the binding energy per nucleon at $T=0$, beeing of the order of a few MeV at $\rho=\rho_0$.\\

To quantify this softening, we can calculate the average speed of sound during the hydrodynamic evolution. We define the average $\left\langle c_s (t) \right\rangle$ at a given time $t$ as the average speed of sound over all fluid cells weighted with the energy density of that cell, where the $c_s$ of every cell can be deduced from the EoS as a function of energy and baryon number density.
The speed of sound $\left\langle c_s (t)\right\rangle$ is then averaged over the whole time evolution, where every time step has the same statistical weight. 

\begin{figure}[t]
\centering
\includegraphics[width=0.5\textwidth]{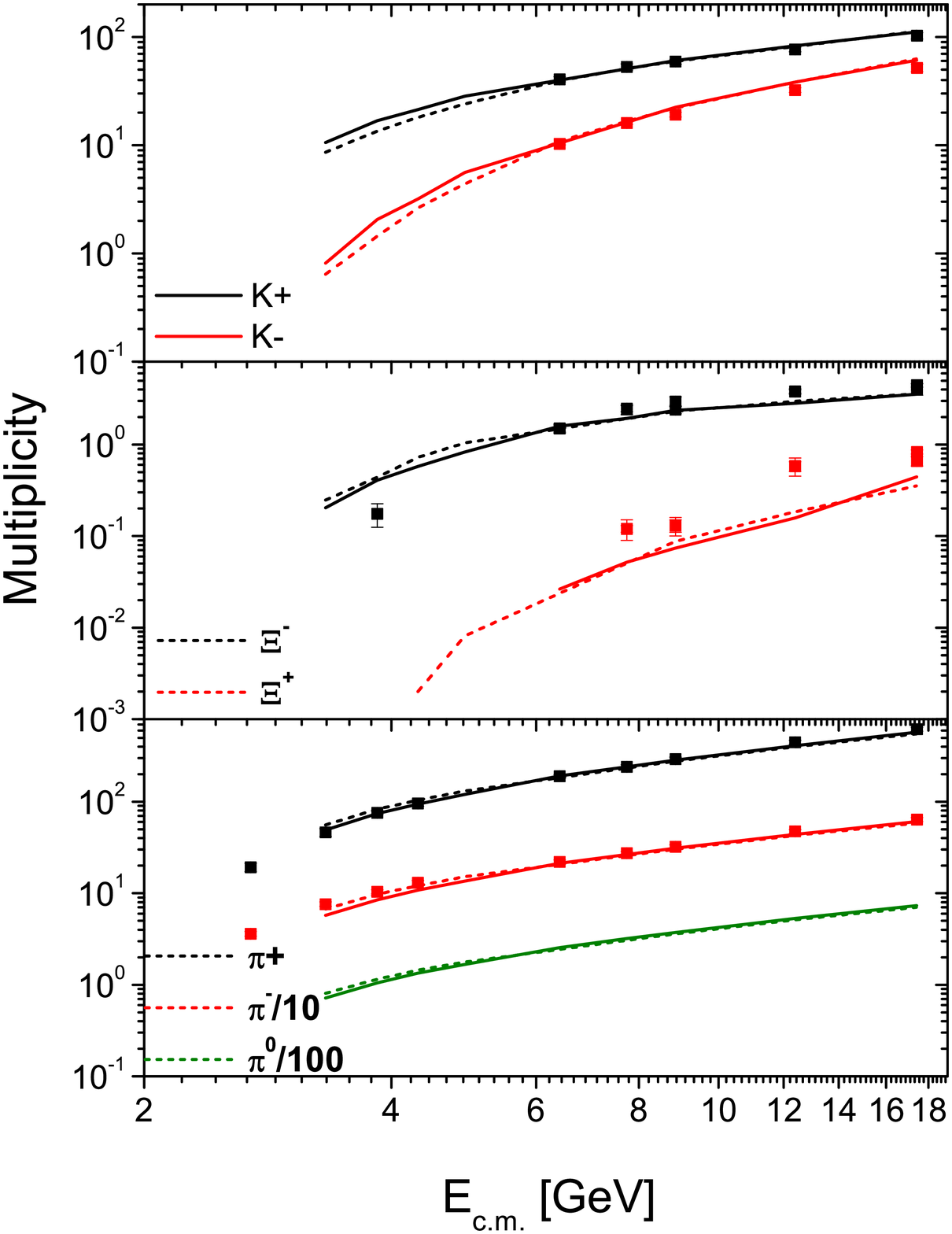}
\caption{\label{multi}(color online)
Total $4\pi$ multiplicities of pions (lower panel), $\Xi$'s (middle panel) and kaons (upper panel). Data \cite{Klay:2003zf,Pinkenburg:2001fj,Chung:2003zr,:2007fe,Afanasiev:2002mx,Anticic:2003ux,Richard:2005rx,Mitrovski:2006js,arXiv:0804.3770,Blume:2004ci,Afanasiev:2002he} are indicated by squares.
}
\end{figure}

\begin{figure}[t]
\centering
\includegraphics[width=0.5\textwidth]{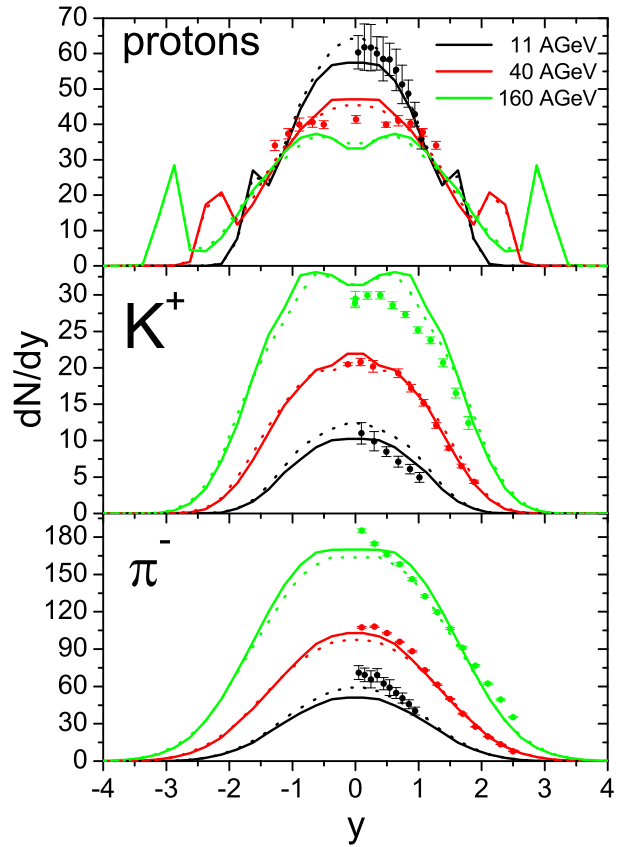}
\caption{\label{dndy}(color online)
Rapidity distributions for different particle species at three different beam energies. Hybrid model calculations, DE (solid line) and HG (dotted lines) results are compared to data (symbols) \cite{Akiba:1996xf,Afanasiev:2002mx}.
}
\end{figure} 

Fig. \ref{cs} shows the excitation function of the averaged speed of sound for both equations of state considered. The hadron gas (blue dashed line) yields higher values of $\left\langle c_s \right\rangle$ than the deconfinement EoS (red solid line). This is expected, since the phase transition leads to a softening of the EoS. Still, in both cases the averaged speed of sound is well above $0.3 \ c$.
It was proposed, that a conical Mach wave created by in medium jets traversing the hot and dense system of a relativistic nuclear collision, could provide the means to experimentally measure the speed of sound in the fireball. Indeed, experiments at the RHIC claim to have observed conical emission in heavy-ion collisions. Applying a 3-particle correlation method, the Mach angle $\theta_M$ was extracted from data \cite{:2008nd}. In a simple Mach cone picture this angle can easily be related to the speed of sound:
\begin{equation}
	\cos(\theta_M)=c_s/v_p,
\end{equation}
where $v_p$ is the velocity of the projectile creating the wave (usually $v_p$ is considered to be close to the speed of light). This simple approximation leads to an estimate for the speed of sound of $\left\langle c_s \right\rangle \approx 0.2  \ c$.\\
Since the partonic jet, which produces the Mach wave, is created in the very early stage of the collision and traverses the medium until freeze-out, the observed angle should also be related to a time average of $c_s$ and not the speed of sound at some specific point in time. Because the systems spends quite a substantial amount of time close to the phase transition 
region, where the EoS is soft, the averaged speed of sound is much lower than the limit for an ultrarelativistic gas $\sqrt{1/3}$.
Still, due to the time and space average, it is substantially larger than the speed of sound in the transition region.
Although the experimental result, obtained at much larger beam energies (and therefore smaller chemical potentials), is not directly related to our results at lower energies, the excitation function of $\left\langle c_s \right\rangle$ shows a saturation at rather moderate energies and, therefore an even lower speed of sound in systems created at RHIC seems unlikely.

\section{Results for final particle properties}

\begin{figure}[t]
\centering
\includegraphics[width=0.5\textwidth]{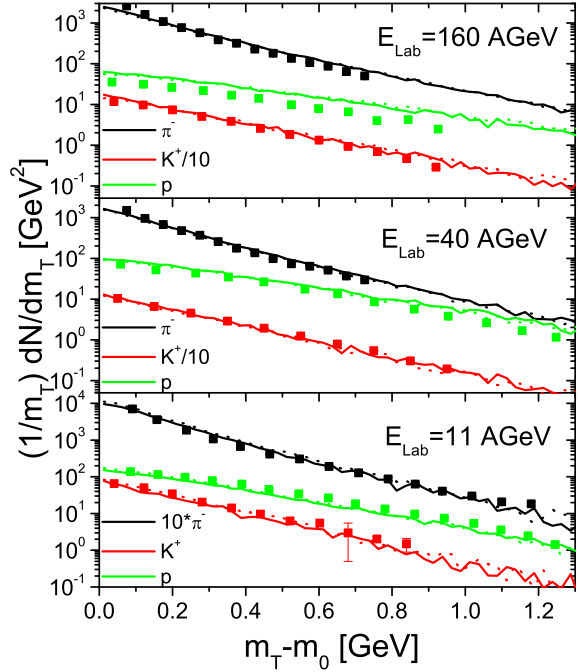}
\caption{\label{dndmt}(color online)
Mean transverse mass spectra of different particle species ($\pi^-$, $K^+$ and protons) at three different beam energies compared to data \cite{Klay:2003zf,Ahle:2000wq,Akiba:1996xf,Afanasiev:2002mx,Alt:2006dk}.
}
\end{figure}

\begin{figure}[t]
\centering
\includegraphics[width=0.5\textwidth]{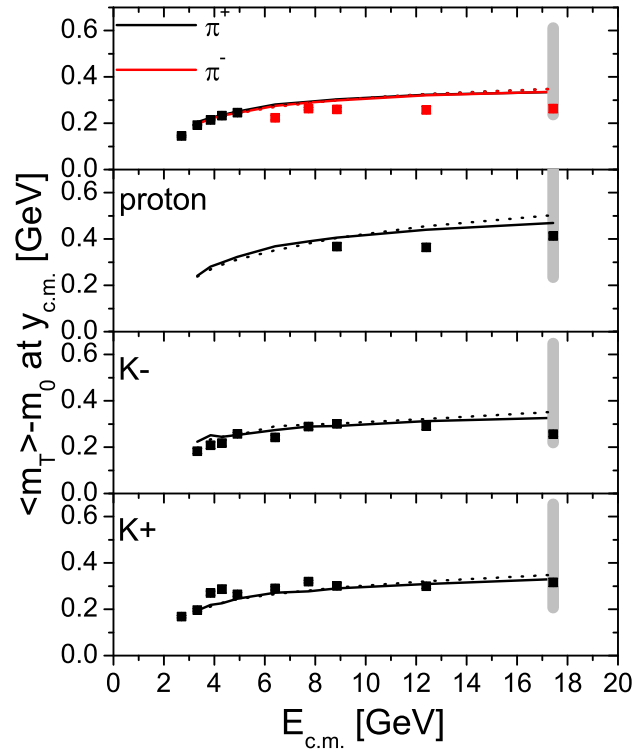}
\caption{\label{mmt}(color online)
Excitation functions of the mean transverse mass of pions (upper panel), protons (upper middle panel), negatively charged kaons (lower middle panel) and positively charged kaons (lower panel) compared to data \cite{Ahle:1999uy,:2007fe,Afanasiev:2002mx}.
}
\end{figure}

In the following, we compare results from our calculations with the deconfinement equation of state (solid lines) including a first order phase transition, to those obtained when an EoS resembling a hadronic resonance gas (HG, dotted lines) is applied to the hydrodynamic evolution.
All results shown are obtained by applying the hybrid model to most central ($b<3.4$ fm) heavy ion reactions (Au+Au/Pb+Pb) in a broad energy range from $E_{\rm lab}=4-160A~$GeV.\\
Note that the observation of particle multiplicity fluctuations and their kurtosis have become the focus of attention concerning the search of the critical endpoint \cite{Stephanov:1999zu,Stephanov:1998dy,Koch:2005vg,Paech:2003fe}. But as has been pointed out in \cite{Schuster:2009jv} fluctuations can be very sensitive on the correct treatment of conserved quantum numbers like baryon number charge and strangeness on an event-by-event basis. It is certainly possible, and planned, to analyze these kind of event-by-event fluctuations in the current hybrid-model. The computational effort for such studies however, is much greater than for bulk observables and we therefore restricted our present study on non event-by-event observables.\\

Fig. \ref{multi} shows the total yields of different particle species as a function of center-of-mass energy of the colliding nuclei.
One can clearly see that the total multiplicities of pions and kaons reproduce the experimental data (squares) reasonably well.
Multistrange hyperons like the $\Xi^-$ are overestimated at the lowest energies, but follow the data very nicely at
energies above $E_{\rm lab} = 11A~$GeV (At lower energies strangeness should be suppressed due to the chemical non-equilibrium of strangeness). This hints to the fact that strangeness is thermalized in heavy ion collisions at energies down to $E_{\rm lab} \approx 11A~$GeV. Only in the sector of heavy anti-baryons (i.e. $\Xi^+$)
the model yields too few particles in the full range of energies investigated. One can clearly see that the excitation functions of total particle multiplicities do not show any signal of the change of the underlying EoS. In contrast, this observable seems much more sensitive to the treatment
of chemical particle freeze out.\\
An alternative and more microscopic way to understand an enhancement in
the production of multi-strange (anti-)baryons is related to the
fragmentation of the initial color flux tubes. While one usually assumes
that the initially produced color strings fragment independently, a high
density of strings may allow for a recombination of color charges at the
string ends, resulting in the formation of a fused string with an
enhanced color field (known as color ropes)
\cite{Biro:1984cf,Knoll:1987gr}. These objects may be seen as an
alternative to QGP formation. Extensive studies within the RQMD model
have shown that particle creation from these strong color fields results
in a substantial enhancement of multi-strange (anti-)baryons \cite{Sorge:1995dp,Sorge:1995vv}.

\begin{figure}[t]
\centering
\includegraphics[width=0.5\textwidth]{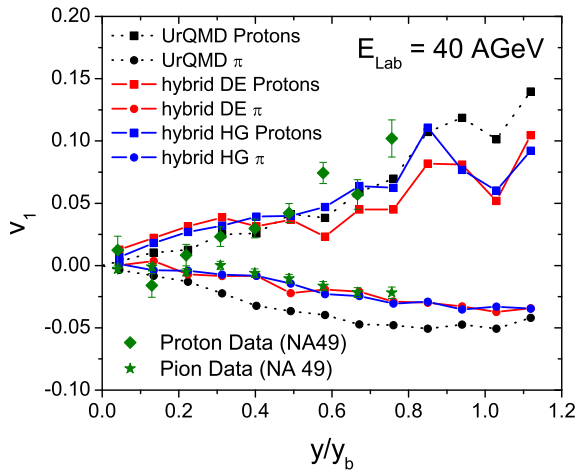}
\caption{\label{v140}(color online)
The directed flow $v_1$ for pions and protons at $E_{\rm lab} = 40 A$ GeV for both hybrid model calculations (blue HG and red DE lines) compared to default UrQMD results (black dashed lines) and data (green stars and diamonds). $y_b$ refers to the beam rapidity \cite{Alt:2003ab}.
}
\end{figure}

\begin{figure}[t]
\centering
\includegraphics[width=0.5\textwidth]{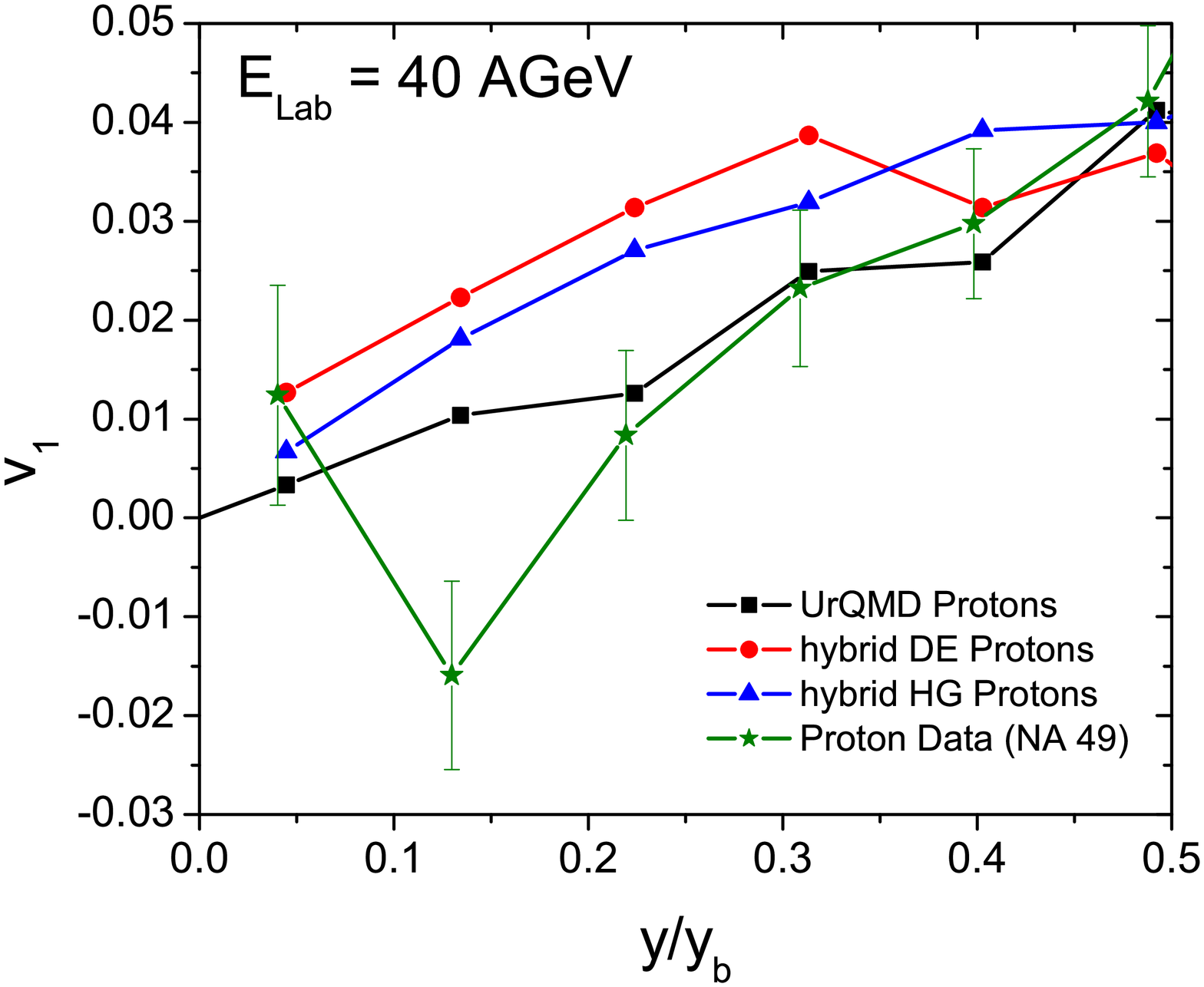}
\caption{\label{v1proton}(color online)
The directed flow $v_1$ for protons at $E_{\rm lab}= 40 A$ GeV for both hybrid model calculations (blue triangles HG and red circles DE) compared to default UrQMD results (black squares) and data (green stars)  \cite{Alt:2003ab}.
}
\end{figure}

Next we turn to the rapidity distributions of different particles (pions, kaons and protons). Fig. \ref{dndy} shows the rapidity distributions for pions, kaons and protons at three different beam energies.
For the lowest energy differences in the proton rapidity spectra can be observed, although both are, within error bars, in agreement with the data.
For the $K^+$ and $\pi^-$ we observe a good agreement up to energies of $E_{\rm lab}=40 A$ GeV, while at the highest energies ($E_{\rm lab} = 160A~$GeV) both hybrid model calculations deviate from experiment. This may be related to our approximate choice of the hypersurface normal $d\sigma_{\mu}^{\rm(1)}$ \footnotetext[1]{ Neglecting space like components of the hypersurface leads to an overestimate of particle production at large rapidities. As we enforce baryon number conservation, the total norm of the rapidity distribution is conserved and therefore less particles should be produced at small rapidities, which can be observed in our results for the highest energies}.

Since the transverse momentum distributions should be sensitive to the transverse dynamics, we next turn to the investigation of the distributions of the transverse mass of different particles (pions, kaons and protons). Comparing the hybrid model calculations to data (squares), in Fig. \ref{dndmt}, we find very good agreement for the lower energies. Especially the pion and kaon spectra are well in line with the data, while the proton distributions deviate slightly.
Comparing the momentum distributions of the different EoS one again observes no noticeable difference between a pure hadron gas EoS and an EoS with a deconfinement transition. This hints to the fact
that the final particle distributions are much more sensitive to
the freeze-out procedure (thermal distribution of particles) and even more to the initial momentum distributions given by UrQMD as compared to the effects of the expansion dynamics.\\
Fig. \ref{mmt} shows the excitation function of the mean transverse mass of different particles,
at midrapidity. Especially for the pions one observes a good agreement with data for low energies. At high energies,
both hydrodynamic calculations overestimate the mean transverse momenta, while the calculation with the DE EoS gives a slightly
better description of the data.

To get an understanding of how much the EoS can (ultimately) influence the final transverse mass spectra, we performed two additional calculations, where we set the pressure (as a function of the energy and baryon-number densities) to either $0$ or $p=1/3 \epsilon$ throughout the whole hydrodynamic evolution. These limits represent the stiffest and softest EoS possible. The results from this comparison at $E_{\rm lab}= 160 A$ GeV are depicted in Fig. \ref{mmt}, as the gray bands, giving an upper ($p=1/3 \epsilon$) and lower ($p=0$) bound on the results from the hybrid model. Note that, especially the pion mean transverse mass can only be accounted for if the (unphysically) soft EoS ($p=0$) is applied. Investigations within the hybrid model, where an EoS including a first order phase transition with a very large latent heat (very small pressure gradients) is employed, come to a similar conclusion \cite{Petersen:2009mz}. In order to explain transverse flow data with ideal fluid dynamics in a hybrid model, an EoS with an unphysically large first order phase transition is needed.\\
The flattening of the mean transverse momentum as a function of beam energy was
assumed to be a signal for a first order phase transition \cite{VanHove:1982vk}. But as has been pointed out above the slow increase of the mean transverse mass at increasing beam energy can be related to non equilibrium effects, and may be modeled by the inclusion of viscous hydrodynamics.\\

Another observable that is deemed to be a signal for the deconfinement phase transition is the so called 'anti-flow' (or third flow component). In \cite{Csernai:1999nf,Brachmann:1999xt,Stoecker:2004qu} it was suggested that a
first-order phase transition leads to a prominent wiggle in the directed flow $v_1=\left\langle p_x/p_T \right\rangle$ as a function of rapidity. Data \cite{Alt:2003ab}, indeed, show a wiggle in the directed flow in most central collisions at $E_{\rm lab}= 40 A$ GeV. Fig. \ref{v140} shows $v_1$ of pions (circles) and protons (squares) as a function of rapidity (over $y_b$, the beam rapidity) for different model calculations, compared to data (green stars and diamonds). Here, we compare both hybrid model calculations (with and without a phase transition)
and a pure transport calculation \cite{Petersen:2006vm} (UrQMD without any hydrodynamic stage). As one can see all models
yield very similar results and none can reproduce the behavior of the proton $v_1$. The pion $v_1$ is very well in
line with data for both hydrodynamic calculations. Note that even the sign change for the pions as compared to
protons is reproduced. Still, changing the EoS does not lead to distinguishable differences in the extracted directed flow.
The directed flow seems to be mostly sensitive to the initial conditions and less to the subsequent hydrodynamical expansion. This is in line with previous findings, where the inclusion of nucleon potentials (mean field effects) in transport models improved the description if the proton directed flow as measured by experiment \cite{Isse:2005nk,Li:2006ez}.

Fig. \ref{v1proton} again shows the directed flow of protons for a smaller window in rapidity. The change of sign in proton flow can be clearly seen in the data, but neither model calculation (hybrid model with and without phase transition) describes this phenomena (not even qualitatively). Disregarding this single data point, the pure transport calculation provides the best description of the directed proton flow. 

\section{Summary}

We discussed the influence of chiral restoration and the deconfinement phase transition in a hybrid model calculation. To model a more realistic equation of state of hot and dense nuclear matter we introduced the Polyakov loop as the order parameter of deconfinement in a chiral hadronic model. This EoS was applied to a micro+macroscopic hybrid model for heavy-ion collisions at intermediate energies. Although the phase transition leads to a softening of the EoS, we find that the average speed of sound in the expanding fireball is of the order of $c_s \approx 0.3 \ c$ over a wide range of energies. This is in contrast to findings at the STAR experiment, where 3-particle correlation data, in connection with a Mach cone scenario, seem to indicate a much smaller average speed of sound.\\
We compared calculations for final particle properties in central A+A collisions over a wide energy range to available data. Here either a hadron resonance gas EoS or the deconfinement EoS was applied. Regarding most observables both models give a reasonable description of data, without tuning any parameters of the model. Two proposed observables for the deconfinement phase transition, the flattening of the mean transverse mass excitation function and the appearance of the third flow component (anti-flow), were investigated within our approach. We showed that the observed flattening of $\left\langle m_T\right\rangle$ cannot be explained by a phase transition, even if a more realistic equation of state is applied. Here other mechanisms, like viscosity and other non equilibrium effects play an essentially more important role.\\
Finally we have shown, that the effect of the EoS on the directed flow is almost completely negligible. No anti-flow is observed in our calculations and especially pion flow is described very well. This leads to the assumption that the proton flow is more sensitive to non-equilibrium effects. Within this picture, the importance of the (non-equilibrium) initial state for flow observables was pointed out.   

\section*{Acknowledgments}
This work was supported by BMBF, GSI and the Hessian LOEWE initiative through the Helmholtz International center for FAIR (HIC for FAIR). The computational resources were provided by the Frankfurt Center for Scientific Computing (CSC).

\end{document}